\RequirePackage{fix-cm}
\documentclass{iopconfser}
\usepackage[utf8]{inputenc}
\usepackage[T1]{fontenc} 
\usepackage{type1cm}    
\usepackage{url}
\usepackage{tabularx}
\usepackage{booktabs}
\usepackage{graphicx}
\usepackage{amsmath}
\usepackage{amsfonts}
\usepackage[hidelinks]{hyperref}
\usepackage{soul}
\usepackage{xcolor}

\begin{document}

\title{Learning-based Hierarchical Tracheal Anatomy Understanding from Sparse Surgical Demonstration Annotations for Ultrasound Robots}

\author{Hiu Ching Cheung$^{1}$, Wenchao Yue$^{1}$, Zhengran Han$^{2}$, Mingcong Chen$^{3}$, Guanglin Cao$^{3}$, Hongbin Liu$^{3}$ and Hongliang Ren$^{1,4}$*}

\affil{$^1$Department of Electronic Engineering, The Chinese University of Hong Kong, Hong Kong, China}
\affil{$^2$Department of Electrical and Electronic Engineering, Imperial College London, London, United Kingdom}
\affil{$^3$ Centre for Artificial Intelligence and Robotics, Hong Kong Institute of Science and Innovation, Chinese Academy of Sciences, Hong Kong, China}
\affil{$^4$ SLAI Shenzhen Loop Area Institute, China}

\email{hiuchingcheung@cuhk.edu.hk, hlren@ee.cuhk.edu.hk}

\begin{abstract}
Tracheostomy requires precise localization of the tracheal incision site; however, conventional manual palpation is subjective and often unreliable, while ultrasound utility remains operator-dependent. This work presents a learning-based framework for hierarchical tracheal anatomy understanding, designed specifically for ultrasound-guided robotic systems. We propose a two-stage perception pipeline integrating a YOLOv8n localization backbone with a sparse, prompt-optimized SAM2 decoder to achieve high-fidelity segmentation from sparse surgical annotations. Our hybrid training strategy, bridging curated laboratory data with unconstrained sequences, ensures clinical robustness. Experimental benchmarks demonstrate that this decoupled architecture effectively balances generalization, precision, and efficiency. The YOLOv8n and SAM2 framework achieves a consistent Mean Dice Similarity Coefficient (DSC) of 0.777 across both controlled and generalized domains. This significantly outperforms U-Net baselines, which often suffer from anatomical fragmentation and performance degradation (Generalization DSC $\le$ 0.494). By constraining mask decoding to targeted, sparse regions of interest, our model achieves a throughput of 6.92 FPS, which is vital for closed-loop robotic teleoperation. This study confirms that a robust hierarchical understanding of tracheal anatomy can be derived by coupling lightweight localization with foundation-scale visual models. Our framework establishes a scalable foundation for standardized, autonomous surgical assistance, effectively navigating the variability of real-world ultrasound to enhance the safety and precision of robotic-assisted tracheostomy.
\end{abstract}
\section{Introduction}
\hspace{1.25em} Tracheostomy is a frequently performed procedure in intensive care units (ICUs) for patients requiring extended mechanical ventilation \cite{frutos2005outcome}. The clinical efficacy and safety of the procedure depend significantly on the accurate identification of the tracheal incision site, typically located between the second and third tracheal rings \cite{xiao2020pilot}. In conventional clinical practice, surgeons often rely on manual palpation to identify anatomical landmarks by sensing the resistance and contours of the tracheal rings. However, this approach is inherently subjective, and its accuracy depends heavily on the operator's clinical experience. In patients with complex neck anatomy, such as those with obesity or edema, manual palpation becomes less reliable due to obscured bony landmarks. Furthermore, this traditional method does not provide quantitative metrics to assess the suitability of the identified site, making it difficult to objectively predict the success rate of the localization or the risk of procedural misalignment \cite{yue2024rasec}.

The ultrasound imaging technique has been integrated into clinical workflows to provide real-time and non-invasive visualization of internal neck structures, offering a more objective alternative to manual palpation. Ultrasound imaging allows for the identification of the trachea and adjacent vascular structures, thereby potentially reducing the incidence of complications \cite{yue2024multimodal}. Nevertheless, the diagnostic utility of ultrasound remains operator-dependent. High-quality B-mode image acquisition requires precise control of the probe's orientation and contact pressure. Suboptimal probe positioning can lead to image artifacts or poor anatomical visibility, which introduces inter-observer variability and may limit the reliability of the diagnostic assessment.

To standardize the imaging process and improve procedural consistency, robotic ultrasound (RUS) systems have been developed. These platforms utilize robotic manipulators to automate probe movement, providing higher precision and repeatability compared to manual handling. Despite these advantages, many current RUS systems are constrained by predefined scanning trajectories or basic force-feedback control. Such systems often lack the capacity for semantic image interpretation \cite{elmekki2025comprehensive}, which limits their ability to autonomously adapt to the anatomical variations present across different patients. Without the integration of high-level perception, robotic systems remain unable to independently identify or quantify the optimal imaging windows necessary for precise incision localization.

\begin{figure}[!t]
    \centering
    \includegraphics[scale = 0.48]{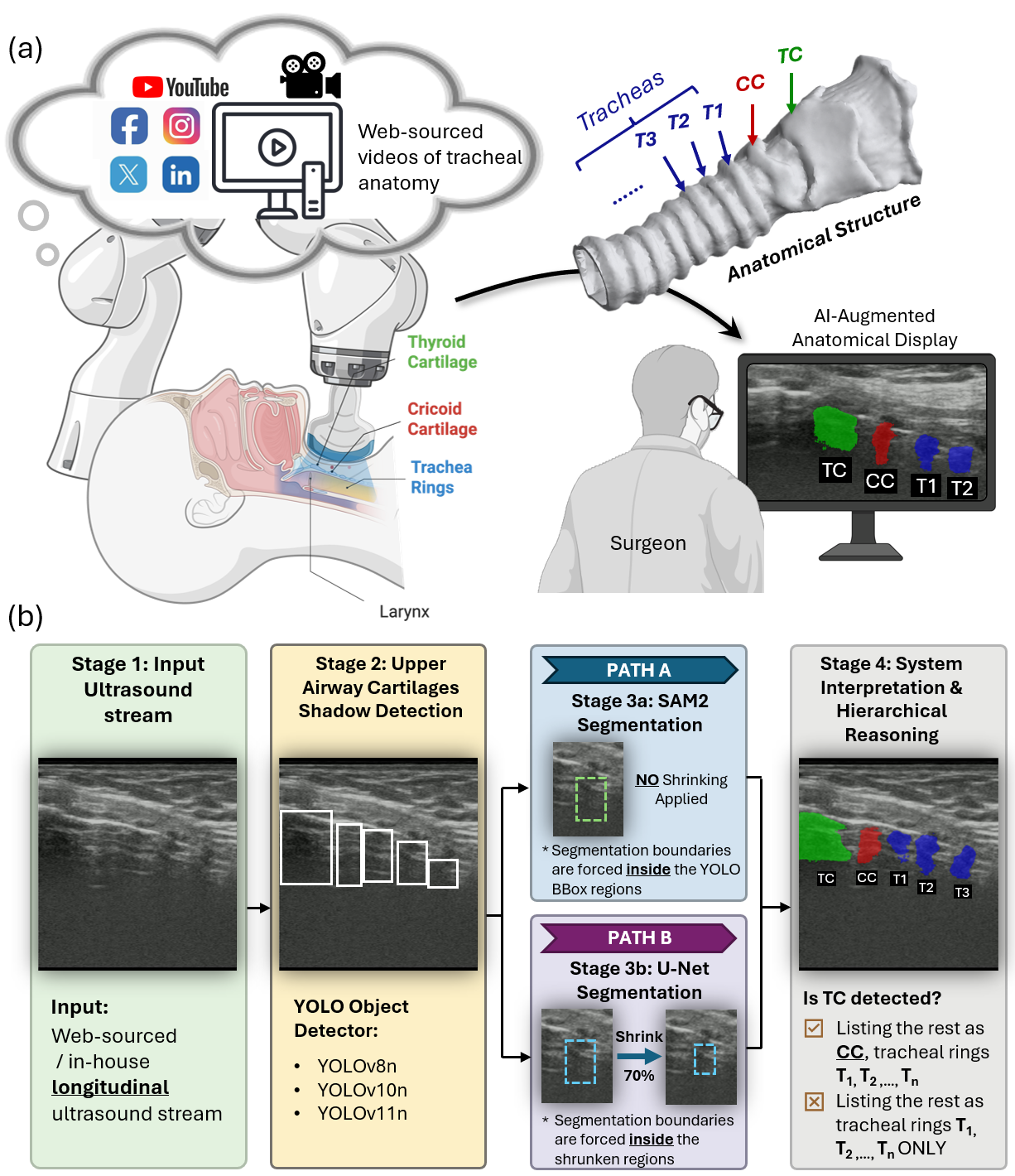}
    \caption{(a) System overview of the proposed framework for hierarchical tracheal anatomy understanding. (b) Detailed architecture of the two-stage detection-guided pipeline.}
    \label{fig::yolo_sam2_unet_framework}
\end{figure}

To address these limitations and enhance robotic autonomy, it is essential to incorporate advanced computer vision for real-time anatomical reasoning. As illustrated in Fig. \ref{fig::yolo_sam2_unet_framework}, we propose a two-stage framework that combines a You Only Look Once (YOLO) object detection backbone with a downstream segmentation engine to achieve robust recognition and high-precision tracking of tracheal structures. In this paradigm, the detector processes the full ultrasound frame to predict localized structural bounding boxes, while the segmentation engine generates fine-grained, pixel-level masks for target regions.

To optimize initial localization, we compare three detection architectures—YOLOv8n \cite{yolov8_ultralytics}, YOLOv10n \cite{wang2024yolov10}, and YOLOv11n \cite{yolo11_ultralytics}—and benchmark their spatial consistency across both controlled and unseen datasets. For pixel-level extraction, we utilize the Segment Anything Model 2 (SAM2) \cite{ravi2024sam}. Its transformer-based streaming memory architecture is particularly well-suited for continuous video processing, as it preserves temporal consistency during transducer movement. Although recent studies demonstrate that integrating the YOLO and SAM architectures can achieve real-time detection with high accuracy \cite{mansoori2025self,verma2026toward, pandey2024validating}, we specifically address the nuances of tracheal ultrasound imagery. Given that ultrasound sequences are characterized by low contrast, acoustic shadowing, and textures that lack standard semantic features, the structural-geometric approach of SAM2 remains the more reliable choice for identifying raw echogenic boundaries.

To validate this approach, we benchmark our hybrid YOLO+SAM2 pipeline against two distinct convolutional baselines: a two-stage YOLO+U-Net \cite{ronneberger2015u} configuration and a standalone U-Net framework. While U-Net \cite{ronneberger2015u} remains a standard for medical image segmentation due to its symmetric encoder-decoder structure and effective local feature preservation \cite{kakade2018identification,neha2024u,asadpour2025artificial,amiri2020fine}, our experiments highlight the vulnerability these architectures exhibit when encountering complex anatomical variations in unseen domains.

Ultimately, this systematic evaluation of decoupled detection-guided segmentation frameworks facilitates a highly robust quantitative analysis of intercartilaginous spaces, enabling the robotic system to localize recommended incision sites with objective confidence. The integration of robust real-time vision algorithms provides a reliable technical basis for more accurate and intelligent robotic-assisted tracheostomy in diverse clinical environments. Our main contributions are summarized as follows:
\begin{itemize}
\item We construct a YOLOv8n-SAM2 integrated robotic ultrasound system for automatic tracheal anatomy recognition, enabling objective, low-latency clinical localization instead of manual operation.
\item Multiple YOLO versions are tested to confirm the optimal YOLOv8n-SAM2 scheme with balanced accuracy and speed, and cross-comparison with U-Net is conducted for technical reference.
\item Our method leverages clinical constraints to differentiate cartilage acoustic shadows and achieves robust hierarchical landmark detection under complex anatomical conditions.
\end{itemize}
\section{Methodology}
\label{sec::method}
\subsection{Cartilage Shadow Tracking and Anatomical Hierarchy}
\label{subsec::cartilage_shadow_detection}
A primary challenge in tracking throat structures via ultrasound is that the thyroid cartilage (TC), cricoid cartilage (CC), and tracheal rings present nearly identical hypoechoic properties. This textural uniformity makes semantic classification based solely on pixel intensity highly ambiguous. To resolve this, we focus on tracking the localized acoustic shadows cast beneath these structures rather than the cartilaginous tissue bodies themselves. These acoustic shadows provide high-contrast, structural boundaries that are inherently more resilient to speckle noise and individual anatomical variation.

To maintain topological consistency and clear representation of the intercartilaginous spaces, data acquisition is restricted to the longitudinal plane. The ultrasound transducer is aligned with the tracheal midline and positioned over the thyroid cartilage. This standardized view exposes a sequential, periodic arrangement of shadows, enabling the autonomous system to measure intercartilaginous gaps and safely identify optimal intervention sites. 

\begin{figure}[t]
    \centering
    \includegraphics[scale = 0.32]{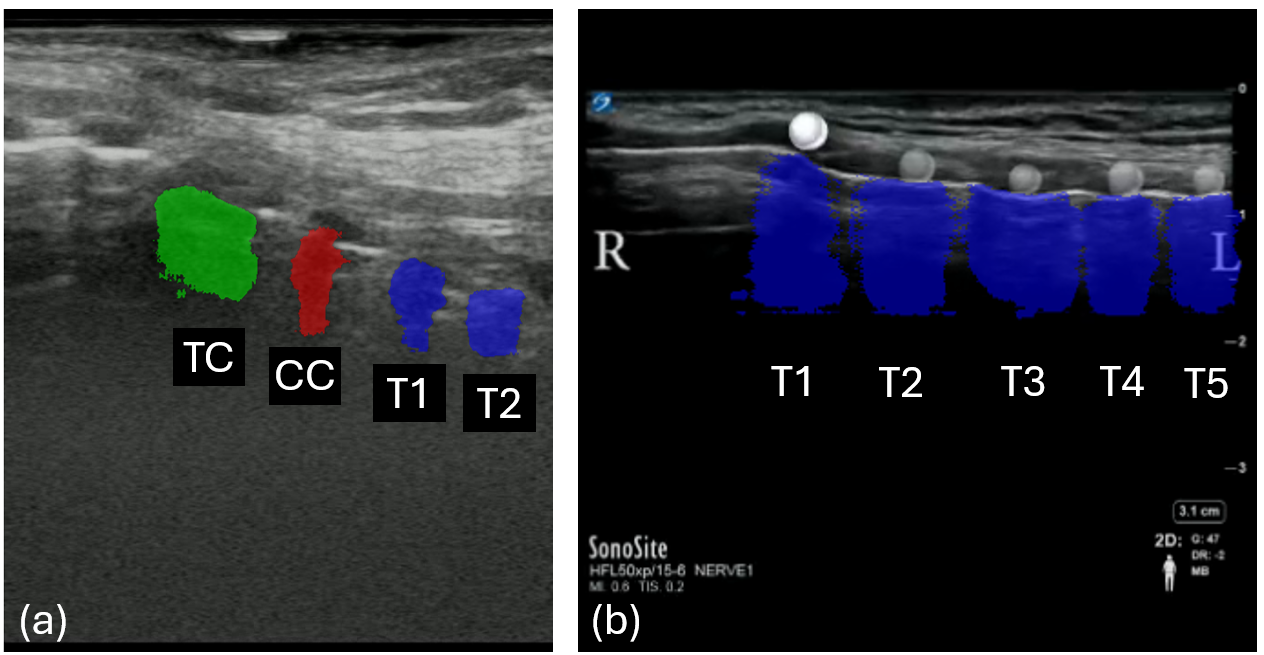}
    \caption{Examples of the hierarchical reasoning logic for upper airway landmark tracking: (a) When the TC is detected, the system verifies its proximity to the frame boundaries and sequentially labels the CC and tracheal rings. (b) If the TC is not detected, the system defaults to labeling the visible acoustic shadows as sequential tracheal rings from left to right.}
    \label{fig::tc_cc_tn_examples}
\end{figure}

By constraining the spatial search domain to this longitudinal orientation, we apply a hierarchical reasoning logic to resolve severe anatomical ambiguities (see Fig. \ref{fig::tc_cc_tn_examples} for a visualization of this process). Under ultrasound, the cricoid cartilage (CC) and subsequent tracheal rings ($T_1 \dots T_n$) exhibit nearly identical hypoechoic properties and acoustic shadow profiles. Since differentiating them purely by visual texture is highly unreliable even for human operators, our upfront object detector simplifies the semantic space into two distinct structural categories: the Thyroid Cartilage ($TC$) and a generalized Cartilage Ring ($CR$) class. The prominent acoustic signature of the $TC$ serves as our absolute global anchor, typically positioned near the peripheral boundaries of the field of view. Once the $TC$ is localized, the system determines its relative boundary proximity to establish a directional tracking vector, retroactively identifying the cricoid cartilage and sequencing the downstream tracheal rings.

To operationalize this logic, let the bounding box predictions generated by the detector be partitioned into a primary anchor $B_{TC} = [x_{TC}, y_{TC}, w_{TC}, h_{TC}]^T$ and a set of $N$ generic cartilage ring proposals $\mathbf{B}_{CR} = \{B_{CR}^1, B_{CR}^2, \dots, B_{CR}^N\}$, where each box coordinate contains its spatial center. Let $W$ define the total pixel width of the frame. The spatial proximity of the global anchor to the frame boundaries determines the searching direction tracking variable, denoted as $D_{side} \in \{\text{left}, \text{right}\}$:

\begin{equation}
D_{side} =
\begin{cases}
\text{left}, & \text{if } x_{TC} < W - x_{TC} \\
\text{right}, & \text{if } x_{TC} \ge W - x_{TC}
\end{cases}
\label{eq:tc_side}
\end{equation}

The generic cartilage ring array $\mathbf{B}_{CR}$ is then sorted by its horizontal centers $x_{CR}$ relative to $D_{side}$ to create an anatomically ordered sequence $\tilde{\mathbf{B}}_{CR} = \{\tilde{B}_{CR}^1, \tilde{B}_{CR}^2, \dots, \tilde{B}_{CR}^N\}$ where $\tilde{x}_{CR}^1$ represents the closest ring adjacent to the $TC$ anchor. The system maps and decodes the final anatomical identities through a directional hierarchy:

\begin{equation}
\text{Anatomical Identity of } \tilde{B}_{CR}^k = 
\begin{cases} 
\begin{cases} 
\text{Cricoid Cartilage (CC)}, & \text{if } k = 1 \\
\text{Tracheal Ring } T_{k-1}, & \text{if } k > 1 
\end{cases}, & \text{if } B_{TC} \text{ is detected} \\
\\
\text{Tracheal Ring } T_k, & \text{if } B_{TC} \text{ is not detected}
\end{cases}
\label{eq:spatial_hierarchy_corrected}
\end{equation}

Conversely, if the primary anchor $B_{TC}$ is not detected or falls outside the field of view, the absolute anatomical baseline becomes temporarily unobservable. In this instance, because the Cricoid Cartilage (CC) cannot be confidently distinguished from surrounding structures without the $TC$ anchor, the system bypasses the CC assignment stage. Under this fallback condition, the framework defaults to a horizontal sort, mapping the visible cartilage ring array ($\tilde{x}_{CR}^1 < \tilde{x}_{CR}^2 < \dots < \tilde{x}_{CR}^N$) directly to an ascending sequence of tracheal rings ($T_1, T_2, \dots, T_n$). To ensure this spatial assumption remains valid, our clinical protocol strictly mandates that the ultrasound transducer orientation marker is consistently aligned toward the patient’s cranial direction; this ensures the left side of the image consistently represents the cranial aspect, thereby rendering the left-to-right sorting logic robust. To maintain visual consistency across all reported results, all anatomical landmarks are rendered with a standardized color schema: Thyroid Cartilage (TC) in green, the Cricoid Cartilage (CC) in red, and sequential tracheal cartilages ($T_1 \dots T_n$) in blue, as illustrated in Fig. \ref{fig::tc_cc_tn_examples}.

\subsection{Standalone Plain U-Net Baseline}
\label{subsec::plain_unet}
To establish a baseline representative of conventional medical image segmentation, we implemented a standalone U-Net architecture \cite{ronneberger2015u}. To optimize this baseline while minimizing peripheral background noise, a localized region-of-interest (ROI) spatial pruning strategy was employed during dataset preparation. Ground-truth coordinate boundaries enclosing the target cartilages were extracted from training sequences and dynamically scaled down to a $70\%$ shrunk spatial domain to tightly bound the high-signal shadow regions. The U-Net was trained exclusively on these tightly cropped regional patches, learning to parse fine structures solely within low-noise environments. 

During inference, however, the model is evaluated on full-frame ultrasound streams without regional cropping or prior localization guidance. As this standalone approach lacks a global mechanism to constrain its search space, it must process the entire uncropped spatial landscape blindly. Consequently, the model exhibits a significant vulnerability to data distribution shifts, leading to tracking instabilities and false-positive mask generation inside peripheral acoustic artifacts.

\subsection{Two-Stage Detection-Guided Frameworks}
\label{subsec::two_stage_frameworks}
As illustrated in Fig. \ref{fig::yolo_sam2_unet_framework}, we implement a decoupled, two-stage segmentation paradigm to resolve the spatial ambiguity and boundary instability inherent in standalone convolutional models. An upfront, high-throughput object detector based on the YOLO architecture enables real-time clinical deployment. The detector processes the full ultrasound frame to predict localized structural bounding boxes. To analyze detection efficiency, we integrate and evaluate three competitive variants: YOLOv8n, YOLOv10n, and YOLOv11n. During inference, the designated YOLO variant screens out peripheral background artifacts and computes the spatial coordinates of the target cartilages, which anchor the downstream segmentation process within a validated anatomical neighborhood.

Using this two-stage constraint, we evaluate two downstream segmentation approaches with distinct spatial input rules:

\begin{enumerate}
    \item YOLO + SAM2: The live, full-resolution ultrasound frame and the unscaled YOLO bounding box coordinates are fed into SAM2. Unlike the U-Net pipeline, SAM2 acts as a promptable foundation model and requires no manual image manipulation, such as spatial shrinking or regional cropping. It consumes raw bounding box coordinates as prompts to decode tissue boundaries. SAM2 utilizes a streaming memory architecture that treats video sequences as a continuous flux rather than disjointed frames, which exploits historical context to resolve fuzzy tissue walls that traditional frame-by-frame models fail to interpret.
    \item YOLO + U-Net: The live frame is dynamically cropped based on the YOLO bounding box predictions, which matches the $70\%$ spatial scaling used during baseline dataset preparation. By restricting the input space to this enclosed territory, the model is protected from full-frame background distractions, which enables stable and localized boundary tracking.
\end{enumerate}

\noindent \textit{Note on Standalone Segmentation:} While SAM2 possesses powerful zero-shot capabilities, executing it as a standalone engine without detection guidance yields inaccurate outputs in noisy medical domains. Thus, we exclude this from our baseline comparisons. Since ultrasound tissue textures lack the distinct semantic boundaries found in natural imagery, a standalone SAM2 fails to isolate individual cartilages from the surrounding throat anatomy. Consequently, the upfront YOLO detector is a strict requirement within our pipeline to anchor the foundation model's prompt mechanism and isolate the target tracheal anatomy.

\subsection{Partial-Label Strategy and Sparse Annotation Protocol}
\label{subsec::partial-label_surgeon_demo}

\begin{table}[htbp]
\centering
\footnotesize 
\setlength{\tabcolsep}{6pt} 
\caption{Dataset Structural Profile and Sparse Ground-Truth (GT) Annotations}
\label{tab:dataset_summary}
\begin{tabular}{@{}lccc@{}}
\toprule
\textbf{Video Sequence} & \textbf{Total Frames} & \textbf{Sampling Interval} & \textbf{Labeled GT Frames (\%)} \\ \midrule
\multicolumn{4}{@{}l}{\textbf{In-house Dataset (Controlled Environment)}} \\
Video 1 & 183 & 20th frame & 10 ($5.46\%$) \\
Video 2 & 70  & 10th frame & 8 ($11.43\%$) \\
Video 3 & 88  & 15th frame & 7 ($7.95\%$) \\
Video 4 & 19  & 4th frame  & 5 ($26.32\%$) \\
Video 5 & 119 & 15th frame & 8 ($6.72\%$) \\
Video 6 & 228 & 25th frame & 10 ($4.39\%$) \\
Video 7 & 38  & 6th frame  & 7 ($18.42\%$) \\ \cmidrule(lr){1-4}
\textit{Subtotal} & \textbf{745} & - & \textbf{55} ($7.38\%$) \\ \midrule
\multicolumn{4}{@{}l}{\textbf{Web-sourced Dataset (Generalization Environment)}} \\
Video a & 68  & 15th frame & 5 ($7.35\%$) \\
Video b & 42  & 8th frame  & 6 ($14.29\%$) \\
Video c & 149 & 20th frame & 8 ($5.37\%$) \\
Video d & 54  & 10th frame & 6 ($11.11\%$) \\
Video e & 85  & 15th frame & 6 ($7.06\%$) \\
Video f & 62  & 20th frame & 4 ($6.45\%$) \\ \cmidrule(lr){1-4}
\textit{Subtotal} & \textbf{460} & - & \textbf{35} ($7.61\%$) \\ \midrule
\textbf{Total Database} & \textbf{1,205} & - & \textbf{90} ($7.47\%$) \\ \bottomrule
\end{tabular}
\end{table}


To optimize data efficiency, we employed a sample-efficient, sparse annotation protocol. As summarized in Table \ref{tab:dataset_summary}, we avoided the labor-intensive task of masking continuous frame sequences through a two-stage curation process. First, we manually cropped raw ultrasound sequences to isolate "keyframe" segments characterized by clear anatomical visualization of the tracheal cartilages and stable transducer motion. Second, we established ground-truth (GT) masks via fixed-interval temporal downsampling. The sampling interval for each cropped sequence was set proportionally to its duration and visual complexity to ensure a representative distribution of labels, with intervals ranging from every 4th to every 25th frame based on the duration of stable anatomical visibility. Out of 1,205 total video frames, only 90 (approximately 7.47\%) required manual labeling. This strategy minimized the overhead of dataset creation while providing sufficient signal context for cross-domain training.

The training data sequence involves a two-stage domain alignment. First, structural profiles and structural variance properties were extracted from a varied set of uncurated web-sourced sequences (Videos a–f). This was followed by highly targeted sparse annotations across both the web-sourced and in-house laboratory datasets (Videos 1–7) to form a complete hybrid training matrix capable of cross-domain validation.

\section{Experiment Validation}
\label{sec::experiment}

\subsection{Experimental Setup and Evaluation Metrics}
To rigorously assess the cross-domain robustness, anatomical tracking accuracy, and computational efficiency of the evaluated frameworks, benchmarks were conducted across two distinct testing environments:
\begin{enumerate}
    \item \textbf{Controlled Domain:} Our curated laboratory-based longitudinal ultrasound dataset featuring standardized imaging configurations and high signal-to-noise ratios. This domain is defined by controlled acquisition parameters, ensuring that the target cartilaginous structures are consistently and clearly identifiable.

    \item \textbf{Generalization Domain:} An unconstrained collection of web-sourced ultrasound sequences \cite{online_video1, online_video2, online_video4, online_video5, online_video6, online_video7}. This set introduces significant domain shifts, including varied acoustic machine settings, probe frequencies, and non-clinical artifacts (e.g., embedded subtitles and background interference) that necessitate robust feature extraction for accurate cartilage identification.
\end{enumerate}

Segmentation accuracy is quantitatively measured using the Dice Similarity Coefficient (DSC). Formally, given a predicted binary segmentation mask $M$ and its corresponding ground-truth annotation $G$, the frame-level spatial overlap is evaluated as:

\begin{equation}
\text{DSC}(M, G) = \frac{2 |M \cap G|}{|M| + |G|} = \frac{2 \cdot \text{TP}}{2 \cdot \text{TP} + \text{FP} + \text{FN}}
\label{eq:dice_coefficient}
\end{equation}
where $\text{TP}$, $\text{FP}$, and $\text{FN}$ denote the pixel-level true positives, false positives, and false negatives calculated across the region of interest, respectively. 

To evaluate temporal performance and suitability for closed-loop robotic teleoperation, the computational footprint is captured via frame-level inference latency. Latency represents the exact high-precision wall-clock time required for a single video frame to propagate through the multi-stage detection and tracking pipeline, captured via hardware timestamps:

\begin{equation}
\text{Latency (ms)} = (T_{\text{post\_inference}} - T_{\text{pre\_inference}}) \times 1000
\label{eq:latency_ms}
\end{equation}

\noindent where $T_{\text{pre\_inference}}$ and $T_{\text{post\_inference}}$ represent the execution start and end times in seconds, respectively. From a clinical deployment standpoint, raw latency dictates system throughput, measured in Frames Per Second (FPS). To accurately reflect continuous, unbottlenecked pipeline execution on sequential data streams, the instantaneous frame rate for an individual frame is defined by the inverse relationship:

\begin{equation}
\text{FPS}_{\text{frame}} = \frac{1000}{\text{Latency (ms)}}
\label{eq:fps_frame}
\end{equation}

To evaluate global performance profiles across whole evaluation sequences within each testing domain, individual frame measurements are aggregated. Due to the non-linear reciprocal relationship between time and rate, system throughput across a sequence of $F$ frames is computed via the harmonic mean of latencies rather than a simple algebraic inversion. 

The aggregated performance metrics for any evaluated model architecture are defined as:

\begin{equation}
\text{Mean DSC} = \frac{1}{N} \sum_{k=1}^{N} \left( \frac{1}{F_k} \sum_{j=1}^{F_k} \text{DSC}(M_j, G_j) \right)
\label{eq:mean_dice}
\end{equation}

\begin{equation}
\text{Mean FPS} = \frac{1}{N} \sum_{k=1}^{N} \left( \frac{1000}{\frac{1}{F_k} \sum_{j=1}^{F_k} \text{Latency}_j} \right)
\label{eq:mean_fps}
\end{equation}

\noindent where $N$ denotes the total number of distinct video sequences within the target evaluation domain (with $N_C = 7$ for the Controlled domain and $N_G = 6$ for the Generalization domain), and $F_k$ represents the number of annotated evaluation frames in the $k$-th video sequence. By calculating the domain-level metrics as the arithmetic average of per-sequence means, we ensure that each video contributes equally to the final performance score, preventing sequences with higher frame counts from disproportionately biasing the results.

It is critical to observe that the calculation of \textit{Mean FPS} in Equation \ref{eq:mean_fps} is subject to Jensen’s Inequality. Because the frame rate function $f(L) = 1000/L$ is strictly convex for $L > 0$, the arithmetic mean of individual frame rates is mathematically guaranteed to be greater than or equal to the reciprocal of the arithmetic mean latency:

\begin{equation}
\mathbb{E}\left[\frac{1000}{L}\right] \geq \frac{1000}{\mathbb{E}[L]}
\label{eq:jensens_inequality}
\end{equation}

Consequently, \textit{Mean FPS} captures the aggregated throughput performance of the pipeline, while the inverse of the mean latency provides a conservative estimate of processing speed, effectively mitigating the biasing effects of temporal outliers.

By explicitly decoupling and balancing both raw latency transformations and their derivative frame rates through Equations \ref{eq:latency_ms} and \ref{eq:mean_fps}, we establish a bounded performance envelope. In medical robotics, processing lag creates a temporal mismatch between physical probe manipulation and the corresponding overlay mask on the display screen. These metrics verify that a network must not only maintain geometric precision across varying target distributions but also sustain a stable visual throughput sufficient to prevent visual tearing or control loop instability during interactive surgical guidance. All experiments were executed on a standardized workstation environment without peripheral model optimizations to guarantee baseline hardware equity.
\subsection{Performance Analysis of YOLO-SAM2 Architectures and Domain Generalization}
\label{subsec::yolo_sam2_comparison_results}

\begin{table}[!t]
\centering
\small 
\setlength{\tabcolsep}{3pt} 
\caption{Evaluation of the proposed YOLO + SAM2 architecture under hybrid and in-house training configurations}
\label{tab:yolo_benchmarks}
\begin{tabular}{@{}lccccc@{}}
\toprule
\textbf{Architecture} & \textbf{Training} & \multicolumn{2}{c}{\textbf{Mean DSC}} & \multicolumn{2}{c}{\textbf{Mean FPS}} \\ 
\cmidrule(lr){3-4} \cmidrule(lr){5-6}
\textit{(Detection + Segmentation)} & \textbf{Dataset} & \textit{Controlled} & \textit{Generalization} & \textit{Controlled} & \textit{Generalization} \\ 
\midrule
\textbf{YOLOv8n+SAM2} & \textbf{Hybrid\textsuperscript{$\dagger$}} & 0.777 & 0.777 & 6.66 & 6.92 \\
YOLOv10n+SAM2 & Hybrid\textsuperscript{$\dagger$} & 0.770 & 0.770 & 6.79 & 6.93 \\
YOLOv11n+SAM2 & Hybrid\textsuperscript{$\dagger$} & 0.752 & 0.775 & 6.74 & 6.88 \\ 
\midrule
YOLOv8n+SAM2 & In-house\textsuperscript{$\star$} & 0.773 & 0.013 & 6.77 & 311.00 \\
YOLOv10n+SAM2 & In-house\textsuperscript{$\star$} & 0.774 & 0.000 & 6.77 & 349.58 \\
YOLOv11n+SAM2 & In-house\textsuperscript{$\star$} & 0.778 & 0.000 & 6.67 & 250.83 \\ 
\bottomrule
\end{tabular}
\vspace{2pt}

\scriptsize 
\raggedright 
\textsuperscript{$\dagger$} Hybrid: Web-sourced (YouTube) and In-house data. \\
\textsuperscript{$\star$} In-house: Custom-captured laboratory data only. 
\end{table}

\begin{figure}[!t]
    \centering
    \includegraphics[scale = 0.32]{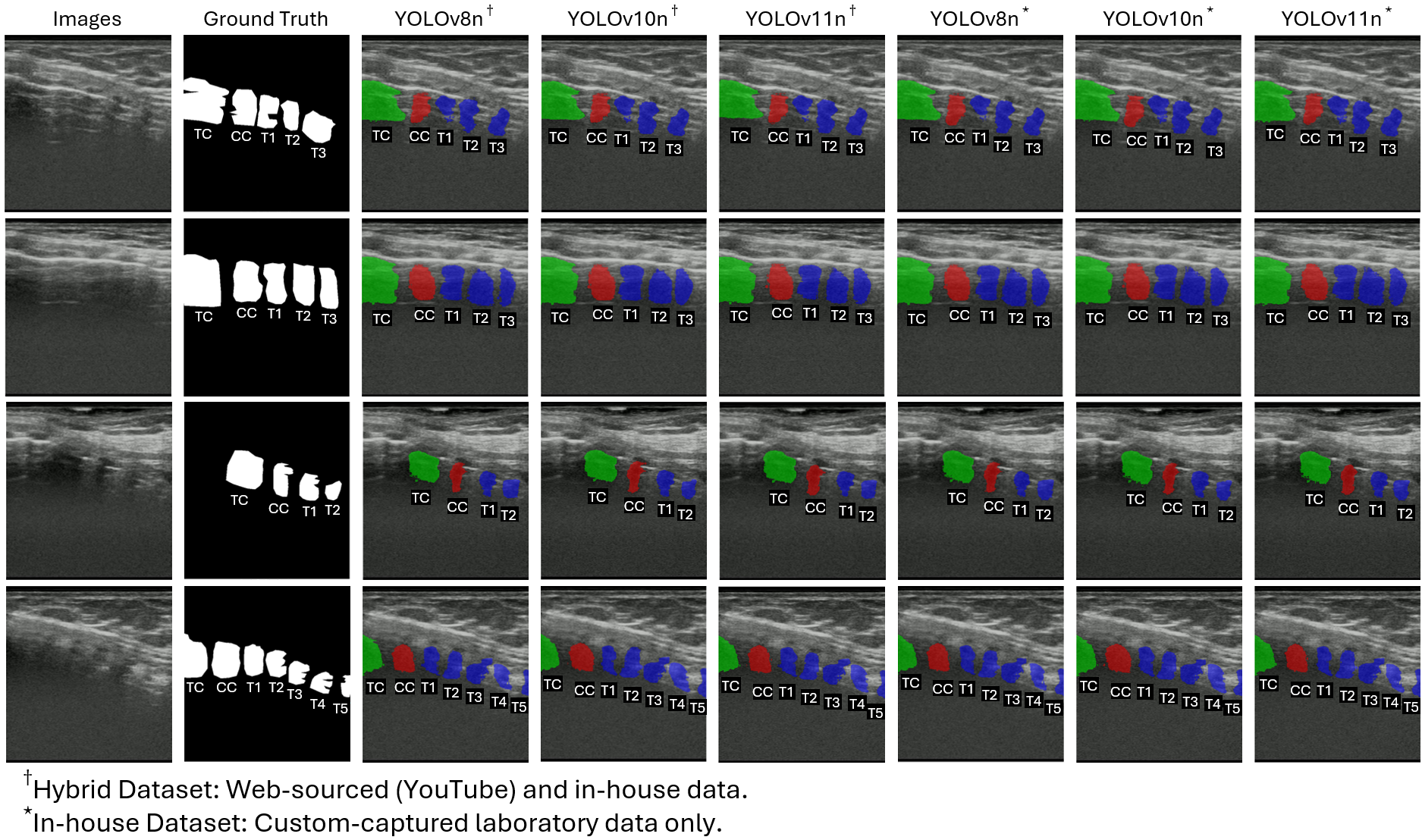}
    \caption{Visualization of segmentation outputs on in-house test frames across different YOLO backbones paired with SAM2, trained on both in-house and hybrid datasets.}
    \label{fig::yolo_sam2_in-house}
\end{figure}

\begin{figure}[!t]
    \centering
    \includegraphics[scale = 0.27]{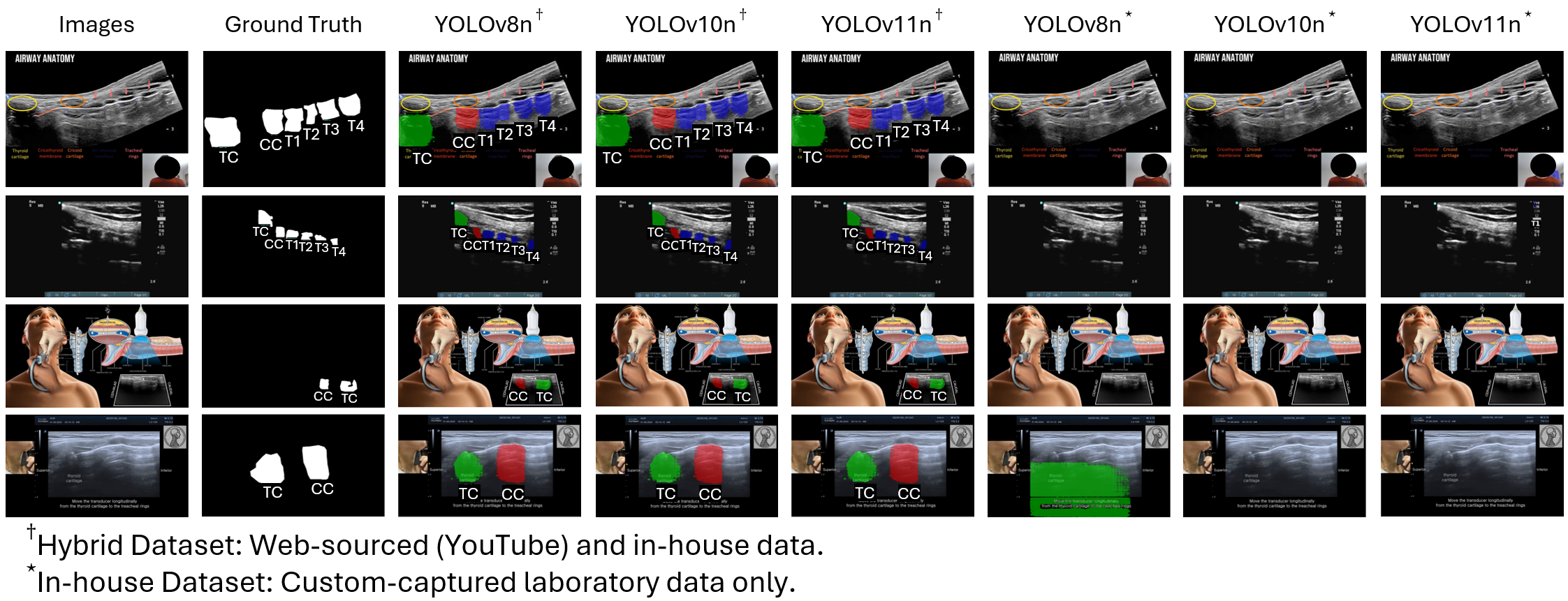}
    \caption{Visualization of segmentation outputs on web-sourced test frames across different YOLO backbones paired with SAM2, trained on both in-house and hybrid datasets. (Note: A black marker over the subject's face was added post-inference solely for publication anonymity; original facial features were preserved during model testing.)}
    \label{fig::yolo_sam2_web-source}
\end{figure}

\begin{figure}[!t]
    \centering
    \includegraphics[scale = 0.31]{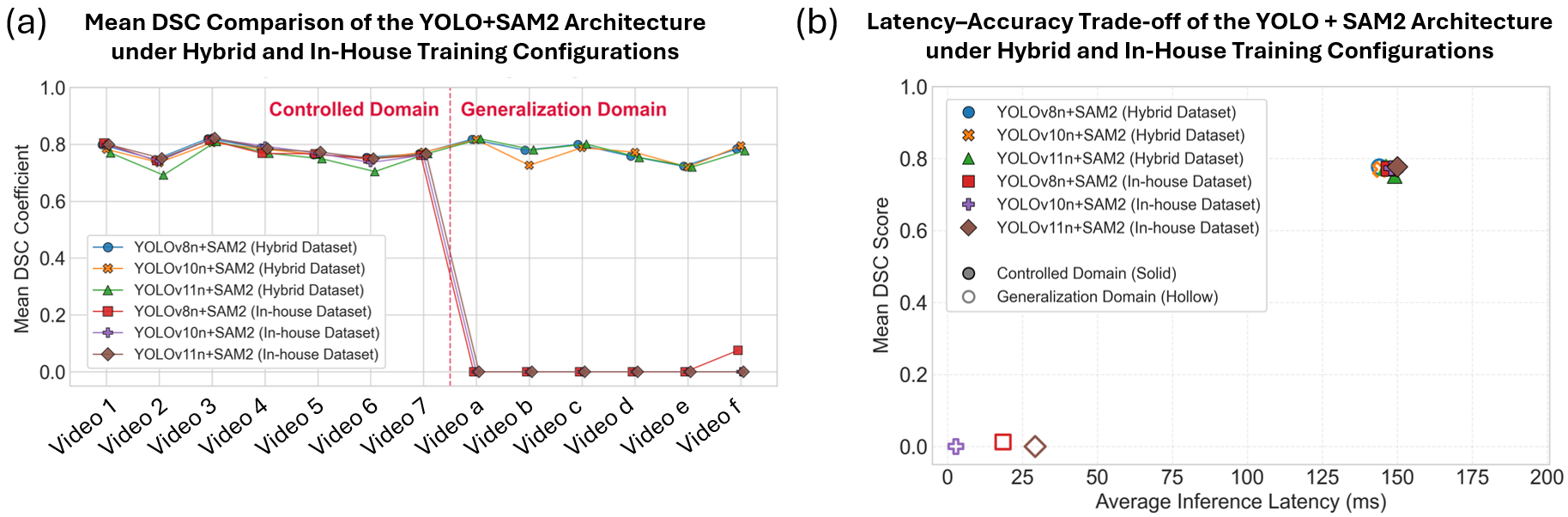}
    \caption{(a) Mean DSC Comparison of the YOLO+SAM2 Architecture under Hybrid and In-House Training Configurations. (b) Latency–Accuracy Trade-off of the YOLO + SAM2 Architecture under Hybrid and In-House Training Configurations.}
    \label{fig::yolo_sam2_mean_dice_latency}
\end{figure}

To determine the optimal configuration for our two-stage pipeline, we systematically evaluated three YOLO backbones—YOLOv8n, YOLOv10n, and YOLOv11n—integrated with SAM2. As detailed in Table \ref{tab:yolo_benchmarks}, our findings confirm that the YOLOv8n+SAM2 architecture, trained on the hybrid dataset, provides the most robust and consistent performance across both test environments.

In the Controlled Domain, all YOLO variants paired with SAM2 exhibit high segmentation accuracy and stable inference speeds, as illustrated in the visual outputs of Fig. \ref{fig::yolo_sam2_in-house}. However, the benefit of our hybrid training strategy becomes evident when evaluating the Generalization Domain. Models trained exclusively on the in-house dataset experience a catastrophic failure when encountering out-of-distribution web-sourced videos, resulting in near-zero Mean DSC scores. As seen in Fig. \ref{fig::yolo_sam2_web-source}, these models fail to identify cartilage structures, often incorrectly mapping artifacts or background noise as target anatomy. Consequently, the abnormally high Mean FPS values reported for these in-house configurations (e.g., $311.00$ to $349.58$ FPS) are systemic anomalies resulting from detection failures, which bypass the downstream segmentation loop entirely; thus, they are excluded from our performance comparison.

In contrast, hybrid training enables the YOLO+SAM2 frameworks to maintain consistent structural recognition across disparate domains. Among these, YOLOv8n+SAM2 achieves the peak Mean DSC of 0.777 in both the controlled and generalization environments. While its Mean FPS ($\sim 6.66\text{--}6.92$) is not the highest, it remains highly competitive and comparable to the other variants. As summarized in the quantitative trends of Fig. \ref{fig::yolo_sam2_mean_dice_latency}, this architecture successfully bridges the generalization gap where in-house training fails.

We conclude that YOLOv8n is the most effective backbone for integration with SAM2. It achieves the optimal balance of high-precision segmentation and reliable generalization, making it the preferred choice for real-time anatomical tracking in diverse clinical settings.
\subsection{Architectural Justification and Computational Profiling}
\label{architectural_justification}
To justify our framework, we compared the YOLOv8n+SAM2 architecture against traditional U-Net baselines. While foundation-scale models like SAM2 are often considered too heavy for real-time deployment, coupling SAM2 with an ultra-lightweight YOLOv8n detector creates an efficient architectural shortcut. The YOLOv8n network acts as a rapid prompt generator, allowing the heavy SAM2 image encoder to skip full-canvas dense embeddings and execute mask decoding exclusively within targeted, sparse ROI coordinates.

Our implementation of the SAM2 memory attention module is specifically optimized for this sparse prompting. By feeding YOLO-generated bounding boxes as queries ($Q_B$), we constrain the attention mechanism to focus exclusively on the anatomical regions of interest. Mathematically, this is expressed as:

\begin{equation}
\text{Attention}(Q_B, K_I, V_I) = \text{softmax}\left(\frac{Q_B K_I^T}{\sqrt{d_k}}\right)V_I
\label{eq:sam2_attention}
\end{equation}

where $Q_B$ is derived from the localized bounding box coordinates, while $K_I$ and $V_I$ denote the spatio-temporal key and value matrices computed from the full resolution visual stream history. This approach bypasses dense, full-canvas attention calculations, effectively bounding the computational overhead within a sparse spatial subset. This architectural choice is the primary driver behind our ability to achieve near-real-time throughput ($\sim 6.66 \text{--} 6.92$ FPS) without sacrificing the temporal consistency offered by the streaming memory module.
\subsection{Comparative Analysis: Proposed Framework versus Conventional Segmentation Baselines}
\label{subsec::yolo_unet_comparison_results}

\begin{table}[!t]
\centering
\small
\setlength{\tabcolsep}{5pt}
\caption{Evaluation of the proposed YOLOv8n + SAM2 architecture against YOLOv8n + U-Net and traditional U-Net baselines}
\label{tab:unet_baselines_comparison}
\begin{tabular}{@{}lccccc@{}}
\toprule
\textbf{Architecture} & \textbf{Training} & \multicolumn{2}{c}{\textbf{Mean DSC}} & \multicolumn{2}{c}{\textbf{Mean FPS}} \\
\cmidrule(lr){3-4} \cmidrule(lr){5-6}
\textit{(Detection + Segmentation)} & \textbf{Dataset} & \textit{Controlled} & \textit{Generalization} & \textit{Controlled} & \textit{Generalization} \\
\midrule
\textbf{YOLOv8n+SAM2} & \textbf{Hybrid\textsuperscript{$\dagger$}} & 0.777 & 0.777 & 6.66 & 6.92 \\
YOLOv8n+U-Net & Hybrid\textsuperscript{$\dagger$} & 0.339 & 0.601 & 1.45 & 2.14 \\
YOLOv8n+U-Net & In-house\textsuperscript{$\star$} & 0.339 & 0.017 & 1.36 & 11.16 \\
U-Net & Hybrid\textsuperscript{$\dagger$} & 0.651 & 0.494 & 7.48 & 7.47 \\
U-Net & In-house\textsuperscript{$\star$} & 0.601 & 0.116 & 7.50 & 7.51 \\
\bottomrule
\end{tabular}
\vspace{2pt}

\scriptsize
\raggedright
\textsuperscript{$\dagger$} Hybrid: Web-sourced (YouTube) and In-house data. \\
\textsuperscript{$\star$} In-house: Custom-captured laboratory data only. 
\end{table}

\begin{figure}[t]
    \centering
    \includegraphics[scale = 0.46]{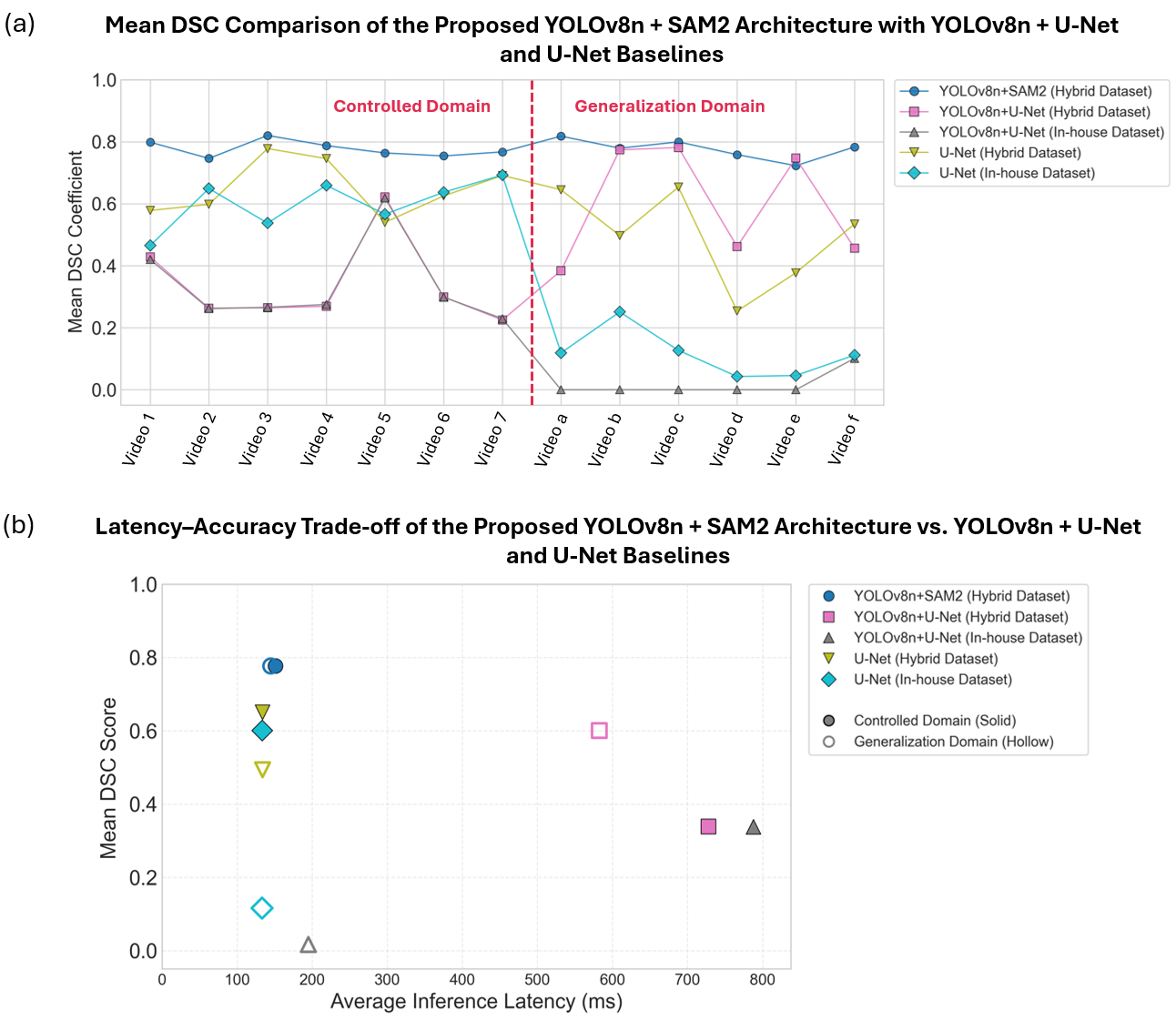}
    \caption{(a) Mean DSC Comparison of the Proposed YOLOv8n + SAM2 Architecture with YOLOv8n + U-Net and U-Net Baselines. (b) Latency–Accuracy Trade-off of the Proposed YOLOv8n + SAM2 Architecture vs. YOLOv8n + U-Net and U-Net Baselines.}
    \label{fig::yolo_unet_mean_dice_latency}
\end{figure}

\begin{figure}[t]
    \centering
    \includegraphics[scale = 0.35]{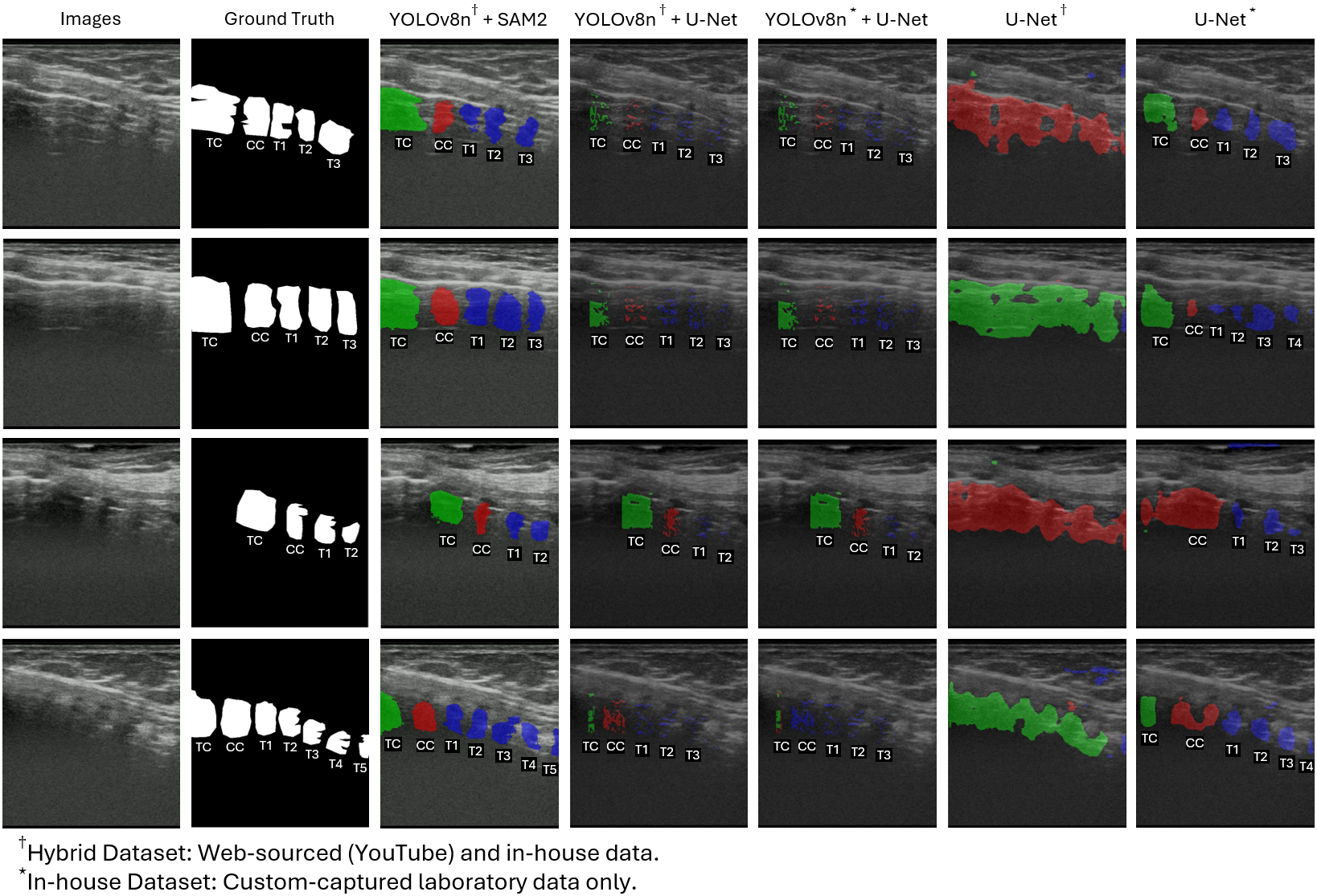}
    \caption{Visualization of segmentation outputs on in-house test frames across different architectural paradigms: YOLO paired with SAM2, YOLO paired with U-Net, and standalone U-Net, trained on both in-house and hybrid datasets.}
    \label{fig::yolo_vs_unet_in-house}
\end{figure}

\begin{figure}[t!]
    \centering
    \includegraphics[scale = 0.33]{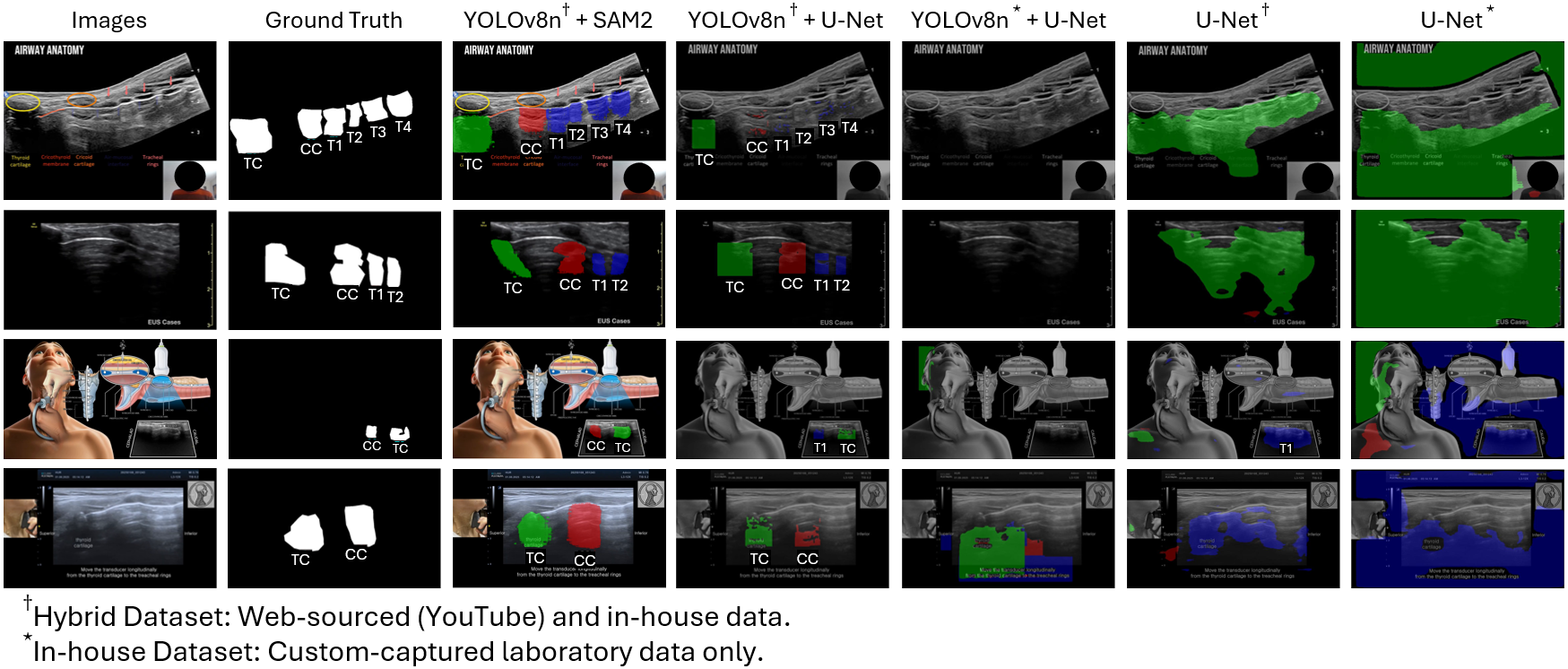}
    \caption{Visualization of segmentation outputs on web-sourced test frames across different architectural paradigms: YOLO paired with SAM2, YOLO paired with U-Net, and standalone U-Net, trained on both in-house and hybrid datasets. (Note: A black marker over the subject's face was added post-inference solely for publication anonymity; original facial features were preserved during model testing.)}
    \label{fig::yolo_vs_unet_online}
\end{figure}

In evaluating our proposed YOLOv8n+SAM2 (Hybrid) framework against conventional segmentation baselines, we demonstrated that raw quantitative metrics often obscure qualitative limitations in clinical performance. While standalone U-Net models exhibited higher throughput and competitive Dice scores in controlled laboratory settings, they failed to maintain anatomical fidelity in the presence of noise. Specifically, U-Net (Hybrid) often merged adjacent tracheal rings into single, indistinct structures, while U-Net (In-house) proved incapable of generalizing to out-of-distribution data, frequently misidentifying background artifacts as target anatomy.

Furthermore, while the YOLOv8n+U-Net (Hybrid) architecture showed improved adaptability, its segmentation outputs were consistently sparse and fragmented, and its pipeline suffered from significant computational bottlenecks caused by dynamic data re-routing. In contrast, our proposed YOLOv8n+SAM2 (Hybrid) model successfully decoupled structural localization from mask decoding. By leveraging a sparse, prompt-guided attention mechanism, it maintained stable, high-precision segmentation ($0.777$ Mean DSC) and near-real-time throughput across both controlled and generalized domains. Ultimately, these results confirm that our framework is the only tested architecture capable of distinguishing individual tracheal cartilages with the reliability required for interactive, robotic-assisted surgical guidance.

\section{Conclusion}
\label{sec::conclusion}
In this work, we introduced a learning-based framework for hierarchical tracheal anatomy understanding, designed specifically for ultrasound-guided robotic systems. By proposing a two-stage pipeline that integrates a YOLOv8n localization backbone with a sparse, prompt-optimized SAM2 decoder, we successfully addressed the challenges of performing high-fidelity anatomical segmentation from sparse surgical demonstration annotations. Our experimental results demonstrate that this decoupled architecture effectively overcomes the inherent trade-offs between generalization, structural precision, and computational efficiency in medical mechatronics. Through a hybrid training strategy that bridges curated laboratory data with unconstrained, out-of-distribution sequences, our framework achieved superior robustness compared to conventional U-Net paradigms; while traditional models often struggle with anatomical fidelity under domain shifts, our approach leveraged structural-geometric cues to maintain high-precision segmentation across diverse clinical scenarios. Ultimately, this study confirms that a hierarchical understanding of tracheal anatomy can be reliably derived from sparse datasets by coupling lightweight localization with foundation-scale visual models. By constraining mask decoding to targeted regions of interest, our framework bypasses the computational bottlenecks of dense segmentation, achieving the near-real-time throughput essential for closed-loop robotic teleoperation.


Despite these advancements, our approach faces certain limitations. While our current processing speed is sufficient for real-time surgical guidance, it operates below the 500 Hz–1 kHz benchmark required to ensure mechanical stability and prevent haptic chatter in closed-loop control systems \cite{laga2025role}. Achieving these frequencies is computationally demanding, as nonlinear mechanics equations must be resolved around 1,000 times per second, particularly when complex mesh topology changes are involved. Our framework currently serves as a high-fidelity perceptual intelligence layer. Furthermore, the robustness of our framework is tied to the current dataset; the overall sample size and the range of patient anatomical variability remain limited, which may affect generalizability across broader clinical populations. 
We intend to address these constraints in future work by decoupling our visual perception module from dedicated high-frequency haptic-feedback loops and diversifying our training data, as these steps are essential to further enhance the safety, precision, and autonomy of robotic-assisted tracheostomy procedures.

\section{Acknowledgment}
\label{sec::acknowledgment}
This work was supported in part by the Ministry of Science and Technology (MOST) of China under Grant numbers 2025YFE0122500 and 2024YFE0216200; the Innovation and Technology Fund (ITF) of the Hong Kong SAR under Grant number MHP/185/24; the Hong Kong Research Grants Council (RGC) General Research Fund (GRF) under Grant numbers 14216022, 14204524, 14203323, and 14206125; the Research Impact Fund (RIF) under Grant number R4020-22; and the RGC Strategic Topics Grant (STG) under Grant number 1/M-405/25-N.

\bibliographystyle{IEEEtran}
\bibliography{IEEEabrv,references}

\end{document}